# BIMA observations of possible microwave background sources[*]

**Lawrence M. Chernin and Douglas Scott**

Department of Astronomy and Center for Particle Astrophysics, University of California, Berkeley, CA 94720, USA

**Abstract.** We present sensitive upper limits on the 90 GHz flux of known radio and infrared sources in regions associated with possible cosmic microwave background fluctuations at 0.5–1 degree scales. Specifically we look at the MAX GUM region and the region of strongest fluctuation in the MSAM scan. None of the known sources can account for the levels of anisotropy seen.

## 1. Introduction

The study of anisotropies in the Cosmic Microwave Background radiation has now reached the point where experiments on a wide range of angular scales are reporting detections (for a recent review see White, Scott & Silk 1994). Although it is clear that primordial fluctuations are now being seen, there is still a possibility that some of the detections may be due to foreground contamination. It is therefore necessary to consider more carefully whether known radio or IR astrophysical sources can contribute to the measured levels of anisotropy.

In particular, two balloon-borne experiments have recently reported detections, which have variously been described as 'low' or 'high', depending on theoretical prejudice. The Millimetre-wave Anisotropy eXperiment (MAX) has observed a region around the star $\gamma$ UMi (GUM) on three separate occasions (Alsop et al. 1992, Gundersen et al. 1993, Devlin et al. 1994), detecting consistent fluctuations each time. Recent follow-up observations conducted in Cambridge (A. Lasenby, private communication) have indicated that a radio source in the region may have a peculiar spectrum that could cause it to contribute to the high frequency MAX results. The other balloon experiment, the Medium Scale Anisotropy Measurement (MSAM) has detected fluctuations that they interpret as primordial, but there were two regions in their scans that had the signature of unresolved sources (Cheng et al. 1994).

In this note, we report sensitive, high resolution, 90 GHz observations of the quasar 1531+722 (Stickel, Kuhr & Fried 1993), which is associated with the MAX GUM region, and 6 IRAS detections near the stronger of the two possible MSAM unresolved sources, MSAM 1492+82.

## 2. Observations

The 90 GHz observations were performed with the Berkeley-Illinois-Maryland Array (BIMA) located in Hat Creek, California. The observations were performed on July 8, 1994 when the array consisted of six 6 m antennas, with the antennas arranged in the most compact configuration. The interferometer has highest sensitivity for point sources and resolves out uniform distributions, so although the beam width is 140″ (FWHM), less than 0.2% of the total flux from 70″ sources can be detected.

*Send offprint requests to*: Lawrence M. Chernin

[*] The BIMA Hatcreek Interferometer is operated under a joint agreement between the University of California, Berkeley, the University of Illinois and the University of Maryland with support from the NSF grant AST-9320238

Each position was observed for 20 minutes, consisting of four five-minute observations. The primary calibrator was 3C345, with an adopted flux of 4.4 Jy. The flux scale is relative to 3C345 and expected to be internally accurate to within 20%. The absolute positional accuracy is a few arcseconds.

## 3. Results

For the MAX GUM region, no source was strongly detected above the rms noise level of 16 mJy. There may be a $3\sigma$ detection from 1531+722, but even this level is down significantly from the $\sim 300$ mJy reported at 15 GHz (A. Lasenby, private communication). Thus 1531+722 probably has a typical synchrotron spectrum, although there may be some deviation at 10–15 GHz. We estimate that this source could contribute at most $\simeq 0.016/7.9 = 0.002$ of the detected signal at 90 GHz (which corresponds closely to the MAX 3.5 cm$^{-1}$ channel). The MAX beam sidelobes may possibly contain emission from outlying IRAS sources such as $15^h32^m26.8^s$ $+71^0 52'43''$ (J2000) which is 14.6' from 1531+722. There are up to 60 IRAS sources within a degree radius of 1531+722 (most are in the reject catalogs).

The sources from the MSAM 1492+82 field were selected based on the criterion of being within 19' of (RA,DEC)=($14^h92, 82°0$). This range is based on the MSAM beam width of 30' (FWHM) and pointing error of a few arcminutes (we use J2000 coordinates which are a few arcminutes offset from the 1993 coordinates). All sources have strong IRAS detections in at least one of the four IRAS bands, but generally do not qualify to be in the point- or faint-source catalogs. There are no known radio sources in this field.

Table 1. Limits on IRAS detections near MSAM 1492+82

| Source | distance from $14^h92 + 82°0$ | 90 GHz rms[1] (mJy) |
|---|---|---|
| IRAS 1452+8155 | 8.'6 | 16 |
| IRAS 1452+8146 | 14.'9 | 21 |
| IRAS 1456+8145 | 15.'0 | 25 |
| IRAS 1502+8154 | 15.'1 | 45 |
| IRAS 1502+8152 | 15.'7 | 23 |
| IRAS 1453+8215 | 19.'0 | 23 |

[1] rms calculated from all pixels within central 50''.

None of these positions showed any obvious signal with the exception of perhaps a $3\sigma$ detection from IRAS 1502+8154 (galaxy: 7ZW 582), which *is* in the IRAS point source catalog. Table 1 shows the 90 GHz rms noise limits for these positions. For the MSAM field, these sources would contribute $\leq 1.5\%$ of the MSAM flux at 90 GHz.

## 4. Conclusions

Since the microwave background experiments are reporting detections at extremely low levels, it is possible that foreground sources, perhaps with unusual spectra, could be affecting the results. It is therefore useful to follow up such detections in all possible ways. We have presented a simple check on anisotropy claims at degree-scales, namely high resolution observations of known point sources in the region. For the recent MAX GUM and MSAM detections we find null results for the associated sources at radio or IR wavelengths. This rules out some possibilities for explaining the anisotropy detections other than through primordial fluctuations.

*Acknowledgements.* We would like to thank Jack Welch for scheduling the BIMA observations and Anthony Lasenby for information on the 1531+722 source.

This article was processed by the author using Springer-Verlag LaTeX A&A style file *L-AA* version 3.